\documentclass[
 aps,
 reprint,
 superscriptaddress,
 amsmath,
 amssymb,
]{revtex4-2}

\usepackage{graphicx}
\usepackage{dcolumn}
\usepackage{bm}
\usepackage{hyperref}
\usepackage{float}
\usepackage{color}
\usepackage{amsmath}
\usepackage{amssymb}
\usepackage{wasysym}
\usepackage{multirow}
\usepackage[normalem]{ulem}
\usepackage{upgreek}
\usepackage{braket}
\usepackage{siunitx}
\usepackage{booktabs}
\usepackage{placeins}

\begin{document}

\preprint{APS/123-QED}


\title{A two-dimensional optomechanical crystal for quantum transduction}

\author{Felix~M. Mayor}
\thanks{These authors contributed equally}
\author{Sultan Malik}
\thanks{These authors contributed equally}
\affiliation{Department of Applied Physics and Ginzton Laboratory, Stanford University, 348 Via Pueblo Mall, Stanford, California 94305, USA}
\author{André~G. Primo}
\thanks{These authors contributed equally}
\affiliation{Department of Applied Physics and Ginzton Laboratory, Stanford University, 348 Via Pueblo Mall, Stanford, California 94305, USA}
\affiliation{Gleb Wataghin Institute of Physics, University of Campinas, 13083-859 Campinas, SP, Brazil}
\author{Samuel Gyger} 
\thanks{These authors contributed equally}
\author{Wentao Jiang}
\affiliation{Department of Applied Physics and Ginzton Laboratory, Stanford University, 348 Via Pueblo Mall, Stanford, California 94305, USA}
\author{Thiago~P. M. Alegre}
\affiliation{Gleb Wataghin Institute of Physics, University of Campinas, 13083-859 Campinas, SP, Brazil}
\author{Amir~H. Safavi-Naeini}
\email{safavi@stanford.edu}
\affiliation{Department of Applied Physics and Ginzton Laboratory, Stanford University, 348 Via Pueblo Mall, Stanford, California 94305, USA}

\date{\today}

\begin{abstract}
Integrated optomechanical systems are one of the leading platforms for manipulating, sensing, and distributing quantum information. The temperature increase due to residual optical absorption sets the ultimate limit on performance for these applications. In this work, we demonstrate a two-dimensional optomechanical crystal geometry, named \textbf{b-dagger}, that alleviates this problem through increased thermal anchoring to the surrounding material. Our mechanical mode operates at \SI{7.4}{\GHz}, well within the operation range of standard cryogenic microwave hardware and piezoelectric transducers. The enhanced thermalization combined with the large optomechanical coupling rates, $g_0/2\pi \approx \SI{880}{\kHz}$, and high optical quality factors, $Q_\text{opt} = 2.4 \times 10^5$, enables the ground-state cooling of the acoustic mode to phononic occupancies as low as $n_\text{m} = 0.35$ from an initial temperature of \SI{3}{\kelvin}, as well as entering the optomechanical strong-coupling regime. Finally, we perform pulsed sideband asymmetry of our devices at a temperature below \SI{10}{\milli\kelvin} and demonstrate ground-state operation ($n_\text{m} < 0.45$) for repetition rates as high as \SI{3}{\MHz}. Our results extend the boundaries of optomechanical system capabilities and establish a robust foundation for the next generation of microwave-to-optical transducers with entanglement rates overcoming the decoherence rates of state-of-the-art superconducting qubits.
\end{abstract}

\maketitle



The integration of mechanical systems with photonic circuits~\cite{Safavi-Naeini.Thourhout.ea:2019} offers a versatile platform for high-performance sensors~\cite{Li.Ou.ea:2021}, signal processing~\cite{Eggleton.Poulton.ea:2019} and exploring macroscopic systems in the quantum regime~\cite{Barzanjeh.Xuereb.ea:2022}.
These optomechanical devices~\cite{Aspelmeyer.Kippenberg.ea:2014a} allow for precise detection of motion using light at and beyond the standard quantum limit~\cite{Mason.Chen.ea:2019}, 
enabling the measurement of
acceleration~\cite{Krause.Winger.ea:2012}, 
displacement~\cite{Eichenfield.Camacho.ea:2009},
mass~\cite{Sansa.Defoort.ea:2020},
forces~\cite{Gavartin.Verlot.ea:2012} at both room and cryogenic temperatures,
and the transduction of quantum information between disparate energy scales~\cite{Andrews.Peterson.ea:2014, Forsch.Stockill.ea:2020}.

Silicon-based optomechanical crystals (OMC)~\cite{Eichenfield.Chan.ea:2009} are particularly promising systems with exceptional photon-phonon cooperativity. 
This is due to their high optical and mechanical quality factors combined with large optomechanical couplings, enabled by their sub-wavelength modal volumes.
Their operation at the quantum level is possible using either milli-Kelvin cryogenic environments or by laser cooling the mechanical modes into their ground state from precooled conditions~\cite{Chan.Alegre.ea:2011}.
This enables the preparation of mechanical quantum states ~\cite{Riedinger.Hong.ea:2016}, optical squeezed states~\cite{Safavi-Naeini.Groblacher.ea:2013}, as well as the demonstration of entanglement between mechanical resonators ~\cite{Riedinger.Wallucks.ea:2018}.
The availability of high-frequency mechanical modes in the \unit{\giga\hertz} regime, combined with piezoelectric~\cite{jiang_optically_2023, Meesala.Wood.ea:2024} or electrostatic coupling~\cite{Zhao.Bozkurt.ea:2023}, make these structures ideal for coherently transducing quantum signals from the microwave to the optical domain to build superconducting qubits-based quantum networks ~\cite{Han.Fu.ea:2021}.

OMCs are typically nanobeams patterned along one-dimension (1D)~\cite{Chan.Safavi-Naeini.ea:2012} ($\approx\SI{500}{\nano\meter}$ in width) that are released from the substrate and co-confine optical and mechanical modes. 
In such devices, heat only dissipates to the surrounding material along the length of the beam. 
The residual heating through linear and non-linear absorption~\cite{Borselli.Johnson.ea:2006, Meenehan.Cohen.ea:2015},
leads to resonance shifts and optical instabilities at room temperature~\cite{Cui.Huang.ea:2021},
while at cryogenic temperatures heating of the mechanical mode hinders operation in the quantum limit.
Therefore various strategies are being pursued to improve thermalization and minimize residual heating. 
First, utilizing non-suspended OMCs may help with thermal anchoring~\cite{Kolvik.Burger.ea:2023} as the heat can propagate into the bulk of the chip. 
Second, cooling the device using a buffer gas, as demonstrated with ${}^3\text{He}$ at $\approx\SI{2}{\kelvin}$~\cite{qiu_laser_2020} has been effective at maintaining low temperatures while laser driving. However, immersion in liquid ${}^4\text{He}$, which transitions to a superfluid below $\SI{2.17}{\kelvin}$, introduces new and interesting superfluid-structure interactions and dynamics which seem to complicate operating the device as a transducer~\cite{Korsch.Fiaschi.ea:2024,sachkou2019coherent}.

\begin{figure*}[!htbp]
\includegraphics[width = 17cm]{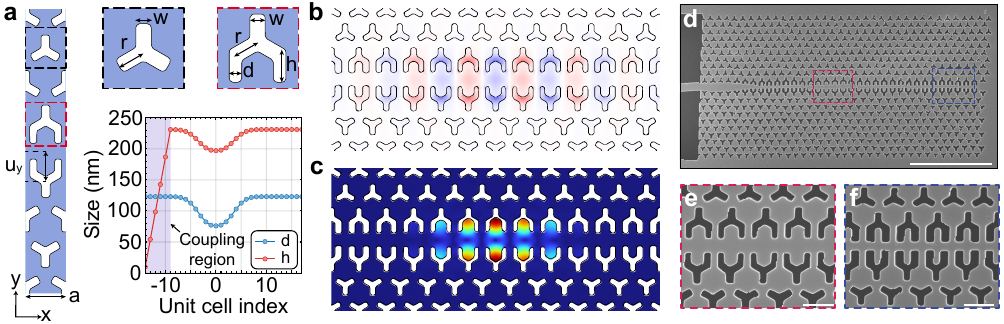}
\caption{\textbf{2D Optomechanical crystal design and implementation.}\small{ \textbf{a}, Schematic of the waveguide unit cell and definition of parameters. Typical parameters are $a = \SI{448}{\nm}$, $w = \SI{93}{\nm}$, $r = \SI{172}{\nm}$ and a fillet radius of \SI{25}{\nm}. The dimensions $d$ and $h$ are modified along the length of the waveguide to confine the optical and acoustic fields. To couple light into the cavity, $h$ is linearly tapered down. \textbf{b, c} FEM simulations of the full cavity displaying the electric field $E_y$ of the fundamental optical mode and the displacement profile of the mechanical mode of interest. \textbf{d},  Scanning electron micrograph of the fabricated optomechanical crystal. Scale bar is \SI{5}{\um}. \textbf{e, f} Close-up of the defect and mirror regions of the waveguide. Scale bars are \SI{500}{\nm}.}}
\label{fig:1}
\end{figure*}


One approach to reducing the challenges posed by heating is to develop two-dimensional (2D) OMCs~\cite{Safavi-Naeini.Hill.ea:2014, ren_two-dimensional_2020, 10.1063/5.0170883}. The initial demonstrations of these devices had limited coupling rates and a complicated mechanical mode spectrum, which largely negated their purported benefits in thermalization. A major advance in 2D OMC design~\cite{ren_two-dimensional_2020} led to photon-phonon coupling rates on par with their 1D counterparts while demonstrating improved thermalization in continuous-wave (CW) operation. Nonetheless, 2D OMCs are yet to be incorporated into optomechanical transducers. Part of the challenge is due to their generally higher microwave operating frequency -- more than $\SI{10}{\GHz}$ vs.\ less than $\SI{5}{\GHz}$ for 1D OMCs. This is because fully connected 2D structures are generally stiffer than 1D devices. Higher operation frequency makes piezoelectric coupling to the mode more difficult, requires a higher optical driving power due to a larger sideband detuning, and leads to a mismatch in frequency with most superconducting qubit systems that tend to operate on the $4$ to $\SI{8}{\GHz}$ range. Secondly, integrated optomechanical transducers have mostly operated as optically pulsed devices to avoid issues related to heating dynamics which kick in on a slower time scale than the pulse duration. It is therefore important to demonstrate the advantages of the 2D OMCs while operating them in optically pulsed mode.


Here we present a novel Si-based two-dimensional OMC with high optical and mechanical quality factors. In contrast to prior demonstrations, the device operates within the typical superconducting qubit frequency band and allows for straightforward integration with a piezoelectric transducer~\cite{chiappina_design_2023, jiang_optically_2023, weaver_integrated_2024}. 
The enhanced thermalization allows us to demonstrate ground-state side-band cooling by CW optical driving starting from a fridge temperature of $\approx\SI{3}{\kelvin}$. Moreover, due to improved power handling, the device operates in a stable manner at sufficiently high power to achieve continuous strong coupling between the optical and mechanical modes -- a first for silicon optomechanical devices to our knowledge.
We demonstrate pulsed operation at $<\SI{10}{\milli\kelvin}$ at repetition rates as large as $\SI{3}{\mega\hertz}$ with the mechanical mode occupation still in the ground state.
Our findings provide a pathway to high-repetition entanglement creation utilizing optomechanical transducers~\cite{Krastanov.Raniwala.ea:2021, Zhong.Wang.ea:2020a, Zhong.Han.ea:2020}.


\section{Results}
\subsection{Two-dimensional C-band optomechanical crystal}

\begin{figure*}[!htbp]
\centering
\includegraphics{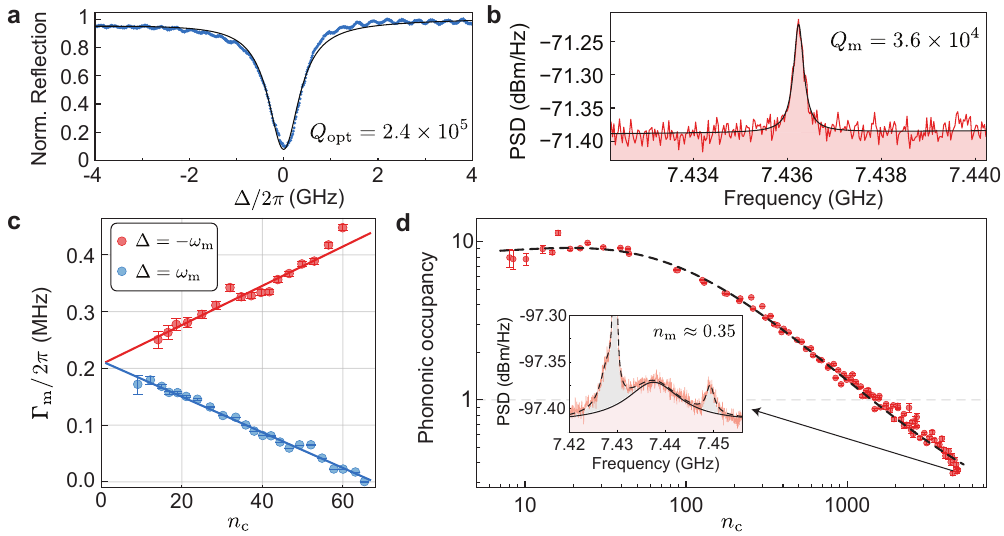}
\caption{\textbf{Ground-state cooling of the mechanical mode.} \small{\textbf{a,} Room temperature optical spectrum as function of the laser-cavity detuning $\Delta = \omega_\text{l} - \omega_\text{c}$. \textbf{b,} Mechanical response of the OMC at $T = \SI{3}{\kelvin}$. \textbf{c,} Backaction cooling and amplification as a function of photon number ($n_\text{c}$), under both red- and blue-detuned excitation. \textbf{d}, Ground state cooling of the mechanical mode. Inset: mechanical spectrum for the highest intracavity occupation in the experiment, $n_\text{c} \approx 4800$.} The error bars represent one standard deviation around the fit value.}
\label{fig:2}
\end{figure*}

Our Si-based two-dimensional OMC is built on a hybrid \textbf{b-dagger} (\textit{boomerang-dagger}) design, where the optomechanical shield is based on the \textit{boomerang} unit-cell (identical to the \textit{blade} cell proposed in Ref.~\cite{aram_optomechanical_2018}) and the defect region is designed in a \textit{dagger} shape.
Fig.~\ref{fig:1}a depicts a schematic of the effective waveguide unit cell of the device together with the parameters defining our design.
We align the $x$-axis of our device along the $[1,0,0]$ crystal orientation of silicon, where the stiffness coefficients are lower resulting in lower mechanical frequencies~\cite{hopcroft_what_2010}.
Optomechanical confinement is achieved by adiabatically modifying the dagger parameters $d$ and $h$ as shown in Fig.~\ref{fig:1}a such as to locally create a density of states inside the photonic and phononic bandgaps (see \ref{Suppl:OMC}).
We use finite element method (FEM) simulations combined with particle swarm optimization to simultaneously optimize the optical and mechanical quality factors, $Q_\text{opt}$ and $Q_\text{m}$, and the single-photon optomechanical coupling rate $g_0$ while minimizing the mechanical frequency $\omega_\text{m}$.
The mode profiles of interest are displayed in Fig.~\ref{fig:1}b and c, showing wavelength-scale confinement in both the optical and acoustic domains.
Our devices operate at an optical frequency $\omega_\text{c}/2\pi \approx \SI{193}{\THz}$, well within the telecom C-band, at $\omega_\text{m}/2\pi \approx \SI{7.5}{\GHz}$, with radiation-limited $Q_\text{opt}>5\times 10^6$ and $Q_\text{m}>1\times 10^9$. 
We compute the optimal $g_0$ accounting for both moving boundary and photoelastic contributions~\cite{primo_quasinormal-mode_2020}, yielding $g_0/2\pi \approx \SI{950}{\kHz}$. The fabricated device (see Methods) is shown in the scanning electron micrograph in Fig.~\ref{fig:1}d with a close-up of the center defect and mirror in Fig.~\ref{fig:1}e and Fig.~\ref{fig:1}f respectively.


We study the devices at room temperature, \SI{3}{\kelvin}, and \SI{10}{\milli\kelvin}. The cryogenic measurements were done at the mixing chamber plate of a dilution refrigerator before and after condensation. We measure the room-temperature optical response using a lensed fiber to send and collect light from an on-chip waveguide butt-coupled to the b-dagger OMC. A typical reflection spectrum is shown in Fig.~\ref{fig:2}\textbf{a} displaying the fundamental optical mode at $\omega_\text{c}/2\pi = \SI{191.7}{\THz}$ ($\SI{1563.5}{\nm}$) as a function of the laser-cavity detuning $\Delta = \omega_\text{l} - \omega_\text{c}$. A Lorentzian fit to the data yields a loaded $Q_\text{opt} = 2.4\times 10^5$ ($2.2\times 10^5$ at \SI{3}{\kelvin}), with an extrinsic coupling rate $\kappa_\text{e}/2\pi \approx \SI{288}{\MHz}$ (measured through coherent sideband spectroscopy), putting the device close to critical coupling. The mechanical spectrum at $T = \SI{3}{\K}$ is assessed through amplitude fluctuations imparted on the reflected optical signal by the thermo-mechanical motion of the OMC. This signal is then measured using a high-speed photodetector and a real-time spectrum analyzer yielding the photocurrent power spectral density (PSD). The mechanical breathing mode of the OMC is found at $\omega_\text{m}/2\pi \approx \SI{7.436}{\GHz}$ with $Q_\text{m} = 3.6\times 10^4$ as shown in Fig.~\ref{fig:2}b.

Our device operates well within the sideband-resolved regime $\omega_\text{m} > \kappa$ ($\kappa/2\pi \approx \SI{800}{\MHz}$). In this regime, the linearized optomechanical interaction under a blue-detuned, $\Delta = \omega_\text{m}$ (red-detuned, $\Delta = - \omega_\text{m}$), coherent excitation is reduced to a two-mode squeezing (beam-splitter) interaction, with Hamiltonian $\hat{H}  = \hbar g (\hat{a} \hat{b}+ \hat{a}^\dagger\hat{b}^\dagger)$ ($\hat{H} = \hbar g (\hat{a} \hat{b}^\dagger+ \hat{a}^\dagger\hat{b})$)~\cite{aspelmeyer_cavity_2014}. This interaction leads to the mechanical mode's effective amplification (cooling), associated with a narrowing (broadening) of its linewidth. Here, $g = g_0 \sqrt{n_\text{c}}$ is the linearized effective optomechanical coupling, enhanced by the number of photons circulating in the cavity, $n_\text{c}$, and $\hat{a}$, $\hat{b}$ denote the annihilation operators for photons and phonons, respectively.

In Fig.~\ref{fig:2}c we characterize the mechanical linewidth, $\Gamma_\text{m}$, as a function of $n_\text{c}$, for both red- ($\Delta < 0$) and blue-detuned ($\Delta > 0$) excitation. In the sideband-resolved regime, $\Gamma_\text{m} \approx \Gamma_\text{m}^0(1+ C)$, where $C = 4 g_0^2 n_\text{c}/(\kappa \Gamma_\text{m}^0)$ is the optomechanical cooperativity and $\Gamma_\text{m}^0$ is the bare mechanical linewidth. A linear fit to the measured data yields $g_0/2\pi = \SI{901 \pm 3}{\kHz}$ ($\Delta = - \omega_\text{m}$) and $g_0/2\pi = \SI{860 \pm 1}{\kHz}$ ($\Delta = \omega_\text{m}$), in good agreement with simulation results. We attribute the discrepancy found for the values of $g_0$ under red and blue excitation to uncertainty in the cavity occupation arising from the Fano lineshape of the optical resonance and variations in the fiber-to-cavity coupling efficiency for different excitation frequencies.

\subsection{Ground-state cooling of the mechanical mode}

A mechanical resonator placed at a temperature $T\ll \hbar\omega/k_\mathrm{B}$, will occupy its ground state nearly perfectly. For a $\SI{7}{\GHz}$ resonator this condition is achieved at the $\SI{10}{\milli\kelvin}$ temperature base temperature of dilution refrigerators. Operating at such low temperatures poses many challenges largely rooted in the extremely small (microwatt-level) available cooling powers, vanishing heat capacity, and small thermal couplings. It is therefore important to consider whether quantum operation is possible at higher temperatures where less exotic cryogenic systems provide orders of magnitude greater cooling power, making scaling to more transducers and control lines more realistic. Higher temperature has been previously considered as a route to improved microwave-to-optical transducers~\cite{Zhong.Wang.ea:2020a,xu_radiative_2020, qiu_laser_2020}. The first demonstration of optical ground state cooling started from a temperature of roughly $T_\text{ini} = \SI{20}{\kelvin}$~\cite{Chan.Alegre.ea:2011}. Rather counter-intuitively, for silicon optomechanical devices, laser cooling from lower starting temperatures becomes \textit{more} challenging, since the temperature increase due to optical absorption heating grows more quickly as the temperature is reduced than the advantage of starting with a lower initial phonon population~\cite{meenehan2014silicon}. Operation in the quantum regime from a lower starting temperature than $\SI{20}{\kelvin}$ is however important for developing transducers, since the superconducting materials used in such devices will need to have a critical temperature $T_\text{c}\gg T_\text{ini}$ to avoid excess quasiparticle losses, and for many common materials $T_\text{c}$'s are in the range of $\SI{10}-\SI{15}{\kelvin}$.   

Starting from a base temperature of $\SI{3}{\kelvin}$, we laser cool our 2D OMC device into its quantum ground state. In Fig.~\ref{fig:2}d, we show the inferred mechanical mode occupation as a function of the laser-driven intracavity photon number $n_\text{c}$. The phonon occupancy, $n_\text{m}$, is obtained through the area under the measured thermo-mechanical response of the acoustic mode at $\SI{7.436}{\GHz}$. In this measurement, we calibrated the photodetection gain for different input powers by intensity modulating the laser at $\SI{7.446}{\GHz}$ and measuring its PSD. The input laser is filtered to mitigate contributions from technical laser noise in the measured mechanical spectrum~\cite{safavi-naeini_laser_2013} reducing the maximum excitation power attainable in the experiment. At low optical powers ($n_\text{c} < 10$), we assume the acoustic mode is thermalized to the plate temperature of $\SI{3}{\kelvin}$, with a phononic occupancy $n_\text{m} = n_\text{th} \approx 7.95$. At high input powers, $n_\text{c} \approx 5000$, the effective temperature of the mechanical mode is cooled to $T_\text{eff}\approx \SI{260}{\milli\kelvin}$, yielding $n_\text{m} \approx 0.35 \pm 0.01$, corresponding to a $74\%$ probability of ground-state occupation. The black dashed line depicts the theoretical prediction for cooling in the presence of a phenomenological increase in the acoustic thermal bath temperature with $n_\text{c}$ (see Methods). In the high intracavity field regime, several modes appear in the mechanical spectra and are included in our fits (see inset of Fig.~\ref{fig:2}d) for an accurate calibration of $n_\text{m}$.

\subsection{Optomechanical strong coupling}

\begin{figure}[!htbp]
\centering
\includegraphics[scale=0.96]{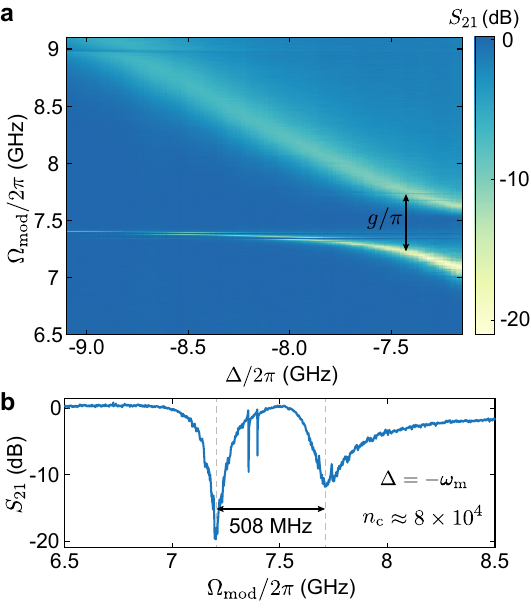}
\caption{\textbf{Optomechanical strong coupling}. \small{\textbf{a,} Measurement of optomechanically-induced transparency (OMIT) as a function of laser detuning $\Delta$. The avoided crossing at $\Omega_\text{mod} \approx -\omega_\text{m}$ results from the strong coupling between the optical and mechanical mode of the OMC. \textbf{b,} Line-cut of \textbf{a} at $\Delta=-\omega_\text{m}$. From the splitting, we infer that we are in the strong coupling regime ($g/\pi=\SI{508}{\mega\hertz}>\kappa/4\pi$). The strong pump field makes additional weakly coupled mechanical modes visible.}}
\label{fig:3}
\end{figure}

Another straightforward consequence of the better thermal anchoring of two-dimensional optomechanical crystals is the strong suppression of thermo-optic nonlinearities. The temperature increase in typical devices red-shifts the optical resonance, leading to an unstable optical response for a sufficiently strong red-detuned pump. For 1D OMCs, this effect usually constrains $n_\text{c} < 2000$~\cite{primo_dissipative_2023, chan_laser_2011, safavi-naeini_electromagnetically_2011} precluding the onset of the strong coupling regime ($4g>\kappa$)~\cite{groblacher_observation_2009}, where the optical and mechanical modes hybridize, hence enabling the coherent state swap between photonic and phononic domains.

To observe strong coupling, we perform optomechanically induced transparency measurements. We intensity-modulate a pump laser detuned by $\Delta$ with respect to the optical mode of the OMC at a frequency $\Omega_\text{mod}$ using an electro-optical modulator and a vector network analyzer. The latter is used to measure the coherent response of the b-dagger OMC, $S_{21} (\Omega_\text{mod})$. When $\Omega_\text{mod} = \omega_\text{m}$, the anti-Stokes sideband arising from the modulation interferes with mechanically-scattered photons giving rise to a transparency window in $|S_{21}|$, indicating a hybridization of mechanical and optical loss channels. We measure this response as a function of the pump detuning $\Delta$ (Fig.~\ref{fig:3}a) for an on-chip pump power of $\SI{9.4}{\milli\watt}$ (corresponding to $n_\text{c} \approx 8\times 10^4$). When the pump is red-detuned by $\Delta = -\omega_\text{m}$, we observe an avoided crossing with a frequency splitting $g/\pi = \SI{508}{\mega\hertz}$ (Fig.~\ref{fig:3}b). From a fit of $S_{21}$ at large detuning $\Delta \approx \SI{-9.5}{\giga\hertz}$, we obtain an optical linewidth $\kappa/2\pi = \SI{870}{\mega\hertz}$. This puts our system well into the strong coupling regime ($4g/\kappa \approx 1.17$), so far elusive in wavelength-scale silicon devices.




\begin{figure*}[!htbp]
\centering
\includegraphics{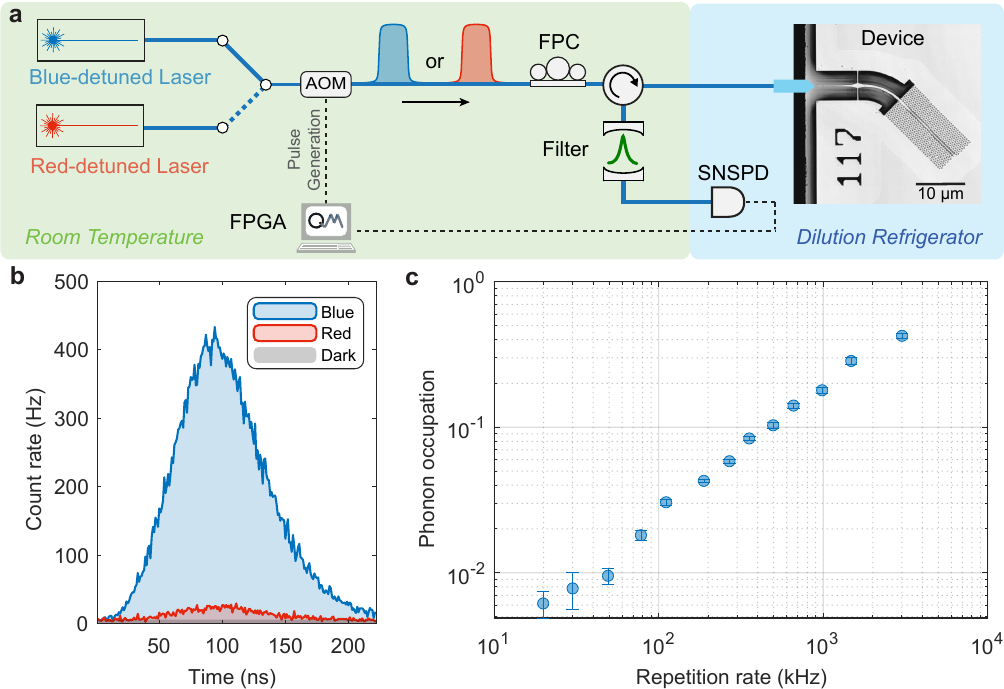}
\caption{\textbf{Pulsed sideband asymmetry measurement.} \small{ \textbf{a,} Simplified setup of the pulsed sideband asymmetry measurement with the device at $T\lesssim \SI{10}{\milli\kelvin}$. Blue- or red-detuned optical pump pulses of length $\tau = \SI{80}{\ns}$ incident on the device generate sideband photons on-resonance with the optical cavity. The reflected light, containing residual pump and sideband photons, is filtered to suppress the pump. The transmitted sideband photons are detected with a single photon detector, and the photon clicks are time-tagged. \textbf{b,} Time trace of the sideband photons generated using blue- and red-detuned pump pulses for a scattering probability of $\approx 5 \%$ and a pulse repetition rate of $\SI{188}{\kHz}$. From the asymmetry between the blue and red count rates, we extract a phonon occupation of $\approx 0.043$.  \textbf{c,} Phonon occupation for a scattering probability of $\approx 5 \%$ for different pulse repetition rates quantifying the effect of optically-induced heating of the mechanical mode. The error bars represent the standard error over multiple sets of measurements (ranging from $4$ to $52$ depending on the repetition rate).}}
\label{fig:4}
\end{figure*}

\subsection{Pulsed sideband thermometry at $T\lesssim \SI{10}{\milli\kelvin}$ }

We characterize a second device at $T\lesssim \SI{10}{\milli\kelvin}$ using pulsed sideband asymmetry measurements on the mechanical breathing mode. Fig. \ref{fig:4}a shows a simplified measurement setup. Starting with a blue- and a red-detuned laser, both parked and locked one mechanical frequency away from the optical cavity resonance, we use two acousto-optic modulators (AOM) in series to generate optical pulses of length $\tau = \SI{80}{\ns}$ with an on-chip peak pump power $P_\text{peak} \approx \SI{7.4}{\uW}$. Through optomechanical interaction, a fraction of the optical pump photons get scattered (or frequency-shifted) into sideband photons resonant with the optical cavity. The reflected light, consisting of residual pump and sideband photons, then passes through a pair of narrowband Fabry-P\'{e}rot cavities (bandwidth $\approx \SI{15}{\MHz}$) that suppress the pump photons with a joint suppression of $>\SI{90}{\dB}$ with respect to the sideband photons. The transmitted sideband photons are then detected by a superconducting nanowire single-photon detector (SNSPD), and the recorded photon clicks are time-tagged. The photon clicks are averaged over multiple trials of the experiment, interleaved every $20$ minutes between the blue- and red-detuned laser. Pulse generation and data acquisition are performed using an FPGA-based real-time controller.

Using the optical pulse parameters and the measured device properties, we estimate the photon-phonon Stokes scattering probability $p_\text{s} \approx  4 g_0^2 n_\text{c} \tau/\kappa \approx 5 \%$, calculated assuming the mechanical mode in the ground-state. Fig.~\ref{fig:4}b shows a time-resolved histogram of the optomechanically scattered sideband photons averaged over millions of experimental runs for blue- and red-detuned lasers for an optical pulse repetition rate of $\SI{188}{\kHz}$. Dark counts arising from spurious light and detector noise are shown in grey ($\approx \SI{5}{\Hz}$). The asymmetry between the sideband count rates under blue-detuned operation (proportional to $p_\text{s} (n_{\text{m}}+1)$) and red-detuned operation (proportional to $p_\text{s} n_{\text{m}}$) provides a direct way to quantify the thermal occupation of our mechanical mode. The asymmetry in count rates stems from the quantum nature of the processes involved: while transitions to higher phonon states induced by the blue-detuned pump are always possible, phonon annihilation due to the interaction with the red-detuned pump is completely suppressed if the mechanical mode is in the ground state. From the asymmetry measured in Fig.~\ref{fig:4}b, we extract a phonon occupation $n_{\text{m}} = 0.043 \pm 0.001$.

The pulsed sideband asymmetry measurements are performed for various optical pulse repetition rates, and the resulting phonon occupation is shown in Fig.~\ref{fig:4}c, revealing the effect of optically-induced heating of the mechanical mode. While the thermal occupation increases with the increasing repetition rate as expected, our device stays in the ground state for a repetition rate as high as $\SI{3.012}{\MHz}$, with a thermal occupation $n_{\text{m}} =  0.42 \pm 0.02$. Our device demonstrates one of the lowest reported thermal occupations for an OMC under pulsed operation, representing a reduction by roughly a factor of eight in the thermal occupation of the mechanical mode compared to 1D structures at similar repetition rates and photon-phonon scattering probabilities~\cite{fiaschi_optomechanical_2021,wallucks_quantum_2020}.


\section{Discussion}
In summary, we designed, fabricated, and demonstrated a proof-of-principle experiment for high-rate phonon-photon pair generation without significant residual heating using a silicon-based 2D optomechanical crystal. 
The enhanced thermal anchoring of our design enables the ground state cooling of the mechanical mode and entering the optomechanical strong coupling regime. 
These demonstrations open the way for full quantum control of integrated optomechanical systems operated at temperatures of $T\approx\SI{3}{\kelvin}$, routinely reached by Gifford-McMahon cryocoolers, allowing for simplified cryogenic infrastructure for quantum sensing and communication applications. Moreover, the large cooperativity that we achieve ($C > 1000$) enables a new class of experiments where photonic states can be coherently swapped into the mechanical domain and vice versa, leveraging nanomechanical resonators as sources of optical quantum states.
Increasing $Q_\text{m}$, which could be achieved by adding more phononic shields to the crystal structure~\cite{doi:10.1126/science.abc7312}, could also propel the use of our mechanical mode as a sensor and as a quantum memory for light. Lastly, our optical quality factors are fabrication-limited and can exceed a few million in similar two-dimensional photonic structures~\cite{Asano:17, MayerAlegre:11}. An enhancement in $Q_\text{opt}$ would reduce the intracavity field required to achieve moderate scattering probabilities and thus further decrease the thermal phononic occupancy under pulsed excitation.

In a next step, our device can be co-integrated with piezoelectric materials to complete a high-rate conversion chain from microwave to optical photons.
For linking two superconducting quantum computing units, the entanglement rate needs to exceed the decoherence rate of local qubits, while at the same time avoiding thermal noise. 
To ensure continuous entanglement availability in superconducting qubit nodes, the demonstrated excitation rates up to $\sim$\unit{\mega\hertz}, considering expected microwave and optical efficiencies~\cite{jiang_optically_2023}, are enough to exceed the $ T_2 = \SI{0.1}{\ms}$ of state of the art superconducting qubits.


\bibliographystyle{naturemag}
\bibliography{references2}

\section*{Methods}

\subsection{Fabrication}

We start device fabrication by using electron-beam lithography (Raith EBPG 5200+, \SI{100}{\kilo\volt}) to pattern the OMCs on a silicon-on-insulator (SOI) chip (\SI{220}{\nm} thick silicon device layer, \SI{3}{\um} thick buried oxide, \SI{725}{\um} silicon handle, Shin-Etsu, $>\SI{3}{\kilo\ohm\cdot\centi\meter}$) with AR-P 62.00 e-beam resist. Next, we do an inductively coupled plasma (ICP) - reactive ion etch (RIE) of the silicon using HBr-based chemistry. The chip is then diced to allow optical fiber access to the coupling waveguide. The substrate is subsequently cleaned in a piranha solution (\SI{96}{\percent} sulfuric acid and \SI{30}{\percent} hydrogen peroxide - \num{3}:\num{1}) to remove any organic residue. Finally, the structure is released in \SI{50}{\percent} hydrofluoric acid (HF) for \SI{2.5}{\minute}. Immediately before loading the device into the cryostat we strip the native oxide with a \SI{2}{\percent} HF dip.

\subsection{Heating model}

The ground-state cooling data shown in Fig.~\ref{fig:2}d is fitted using a model accounting for the presence of pump-induced heating through

\begin{equation}
    n_\text{m} \left( n_\text{c} \right) = \frac{1}{1+C}\left( n^{0}_\text{th}+\frac{\alpha_\text{sat}}{1+\beta_\text{sat} n_\text{c}} n_\text{c}+\alpha_\text{lin} n_\text{c} \right),
\end{equation}
where both saturable and linear absorption terms are used to phenomenologically describe the increase in the temperature of the mechanical mode's thermal bath. The fit results are $\alpha_\text{sat} = 0.324$, $\beta_\text{sat} = 0.019$, and $\alpha_\text{lin} = 0.003$, indicating an increase in the bath's temperature to approximately $\SI{7}{\kelvin}$ (from an initial $T = \SI{3}{\kelvin}$) at $n_\text{c} \approx 5000$.

\section*{Data availability}
The data that support the findings of this study are available from the corresponding author upon reasonable request.

\section*{Acknowledgment}
The authors thank O.~A.~Hitchcock, M.~P.~Maksymowych, R.~G.~Gruenke and K.~K.~S.~Multani for helpful discussions and technical assistance.
This work was primarily supported by the US Army Research Office (ARO)/Laboratory for Physical Sciences (LPS) Modular Quantum Gates (ModQ) program (Grant No. W911NF-23-1-0254).	
Some of this work was funded by the US Department of Energy through grant no.\ DE-AC02-76SF00515 and via the Q-NEXT Center, and by the National Science Foundation CAREER award no.\ ECCS-1941826. 
We also thank Amazon Web Services Inc.\ for their financial support.
Device fabrication was performed at the Stanford Nano Shared Facilities (SNSF) and the Stanford Nanofabrication Facility (SNF), supported by NSF award ECCS-2026822. 
A.G.P and T.P.M.A. acknowledge the S\~{a}o Paulo Research Foundation (FAPESP) through grants 2023/00058-0, 19/09738-9, 18/15577-5, 18/25339-4, and Coordena\c{c}\~{a}o de Aperfei\c{c}oamento de Pessoal de N\'{i}vel Superior - Brasil (CAPES) (Finance Code 001).
S.G. acknowledges the Knut and Alice Wallenberg foundation (grant no.\ KAW 2021-0341).

\section*{Competing interests}
A.H.S.-N. is an Amazon Scholar. The other authors declare no competing interests.
\clearpage
\newpage
\onecolumngrid

\section*{Supplementary information}

\renewcommand{\figurename}{}
\renewcommand{\thesection}{Supplementary Note \arabic{section}}
\setcounter{section}{0}
\renewcommand{\thefigure}{Supplementary Figure \arabic{figure}}
\setcounter{figure}{0}

\renewcommand{\tablename}{Supplementary Data Table}
\setcounter{table}{0}

\section{Measured device parameters}
The data presented in this work was acquired from two devices from different fabrication runs.
Device A (D120) was measured at \SI{3}{\kelvin} after Device B's (D122) thermometry measurements at less than \SI{10}{\milli\kelvin} were cut short due to infrastructure difficulties.

\begin{table}[!htbp]
\caption{\label{tab:parameters}\textbf{Device parameters}}
\begin{tabular}{@{}cSSc@{}}
\toprule
Parameter & {Device A (D120)} & {Device B (D122)} & Method \\
\midrule
$\omega_\text{o}/2\pi$ & \SI{191.7}{\tera\hertz} & \SI{193.9}{\tera\hertz} & Laser wavelength sweep\\
$\kappa_\text{o}/2\pi$ & \SI{0.8}{\giga\hertz} & \SI{1.1}{\giga\hertz} & Laser wavelength sweep\\
$\kappa_\text{o,e}/2\pi $ & \SI{288}{\mega\hertz} & \SI{196}{\mega\hertz} & OMIT\\
$g_\text{o}/2\pi$ & \SI{860}{\kilo\hertz} & \SI{889}{\kilo\hertz} & Mechanical response ($\Delta = \omega_\text{m}$)\\
$\omega_\text{m}/2\pi$ & \SI{7.436}{\giga\hertz} & \SI{7.259}{\giga\hertz} & OMIT\\
$\gamma_i/2\pi$ & \SI{206}{\kilo\hertz} & \SI{715}{\kilo\hertz} &  Mechanical response ($\Delta = \omega_\text{m}$) \\
Fiber to chip coupling efficiency & 43.0\% & 51.1\% & Power meter \\
\bottomrule
\end{tabular}
\end{table}

\section{Optomechanical crystal design}

\begin{figure*}[!htbp]
\centering
\includegraphics[width = 17cm]{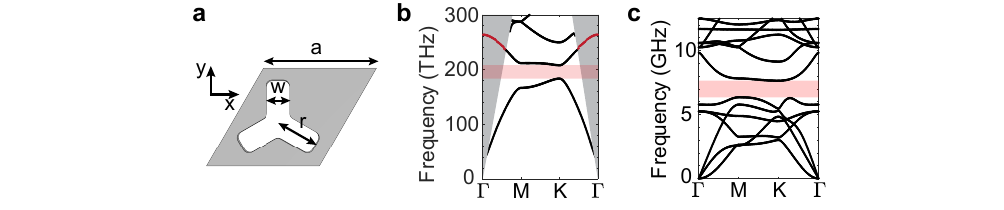}
\caption{\textbf{Optomechanical shield design.} \textbf{a}, Boomerang unit cell schematic. \textbf{b}, Photonic band structure for the even parity modes. The bandgap is highlighted in pink and the light cone in grey. Notice how the bands for the leaky modes (red) inside the light cone do not close the bandgap. \textbf{c}, Phononic band structure. The bandgap is highlighted in pink.}
\label{Suppl:design}
\end{figure*}

To design the optomechanical shield, we consider multiple factors. It should simultaneously have a large optical bandgap at telecom frequencies and a complete phononic bandgap at frequencies commonly used for superconducting qubits. Ideally, we would also like to have the volume of the unit cell be filled with as much silicon as possible, i.e. we would like a large filling factor. This is to help with thermalization and reduce the number of surfaces that can lead to optical scattering. While the snowflake unit cell \cite{safavi2010design} fulfills most of these requirements, OMCs using the snowflake have so far been restricted to mechanical frequencies of $\sim \SI{10}{\GHz}$ \cite{ren_two-dimensional_2020}. This makes it challenging to design and fabricate a transducer that efficiently couples the breathing mode of such an OMC to superconducting circuits. This is mainly due to the smaller required feature sizes of a piezoelectric transducer at higher frequencies. To address this issue, we use the boomerang unit cell \cite{aram_optomechanical_2018} which supports a lower frequency acoustic bandgap than the snowflake at the cost of a smaller relative photonic and phononic bandgap. A schematic of the unit cell is shown in \ref{Suppl:design}a. Finite-element method (FEM) simulations with periodic boundary conditions (\ref{Suppl:design}b) show that for $\SI{220}{\nano\meter}$ thick silicon and $(a,r,w) = (\SI{448}{\nano\meter},\SI{172}{\nano\meter},\SI{93}{\nano\meter})$ the optomechanical shield supports an optical pseudo-bandgap for TE-like guided optical waves. This bandgap spans the telecom frequencies $183-\SI{208}{\tera\hertz}$ and, importantly, does not close inside the light cone \cite{safavi2010design}. Additionally, FEM simulations assuming mechanically-isotropic silicon (\ref{Suppl:design}c) reveal a $\SI{1.30}{\GHz}$-wide complete acoustic bandgap centered at $\SI{6.99}{\GHz}$. The filling factor of the boomerang unit cell is $\SI{74.5}{\percent}$, \SI{15}{\percent} larger than the $\SI{64.1}{\percent}$ filling factor of the snowflake unit cell.

As described in the main text, the dagger parameters in each waveguide unit cell are adiabatically transitioned from the center defect (index 0) to the most external cell (index $N$, for an OMC with $2N+1$ unit cells). The function that parametrizes these variations is given by

\begin{equation}
    v_n = v_{N}-\left(v_{N}-v_\text{0}\right)2^{-\left(\frac{n}{\delta_x}\right)^{M}},
\end{equation}
where $v_n$ is a given parameter of the dagger ($d$ or $h$) for the $n$-th unit cell. The parameters $M$ and $\delta_x$ control how smooth the variation in $v$ is as a function of $n$ and how many waveguide cells will effectively make the transition from $v_0$ to $v_N$.

\begin{figure*}[!htbp]
\centering
\includegraphics[width = 17cm]{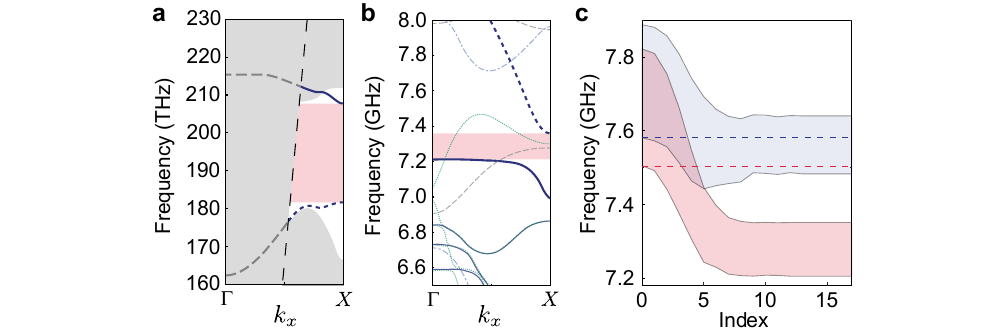}
\caption{\textbf{Effective waveguide band structure.} FEM simulations of the waveguide unit cell of device B displaying its \textbf{a}, photonic and \textbf{b}, acoustic band structures. Solid dark blue bands represent modes with symmetry groups where $\sigma_z = \sigma_y = 1$, meaning they are symmetric for reflections across the $y$ and $z$ axes. Dotted green curves are $\sigma_z = - \sigma_y = 1$, dashed grey curves are $\sigma_z = \sigma_y = -1$, and light blue dash-dotted curves denote $ - \sigma_z = \sigma_y = 1$. Dark blue dashed curves indicate the bands giving rise to the confined optical and acoustic modes.   Results for device A are qualitatively similar. \textbf{c}, Mechanical bandgap as a function of cell index. The blue (red) shaded region shows results for device A (B).}
\label{Suppl:OMC}
\end{figure*}

A FEM simulation of the waveguide unit cell shown in Fig.~\ref{fig:1}a is performed assuming Floquet periodic conditions. It shows the existence of an incomplete (complete) bandgap in the photonic (phononic) domains, as shown in red in~\ref{Suppl:OMC}a~(b). Importantly, the acoustic bands inside the bandgap display different symmetries than our mode of interest minimizing their coupling, as easily verified through the absence of anti-crossings in the red-shaded area.

\ref{Suppl:OMC}c shows the bandgap -- obtained from the same simulations described above -- as a function of the unit cell index. Although the mechanical frequency falls outside the bandgap after index 4 in device B, it keeps a high mechanical quality factor in simulations $Q_\text{m} > 5\times 10^6$, whereas device A has a $Q_\text{m} > 10^9$. In practice, fabrication imperfections allied to the small acoustic bandgaps give rise to a radiation-limited measured $Q_\text{m}$. Adding a surrounding array of rectangular phononic shields could boost our $Q_\text{m}$, at the cost of potentially worse thermal anchoring to the substrate, becoming a less favorable configuration for transduction applications.

All design parameters of Device A and B parameters are provided in Table~\ref{tab:design-parameters}.

\begin{table*}[!htbp]
\caption{\label{tab:design-parameters}\textbf{Design parameters}}
\begin{tabular}{@{}cSSc@{}}
\toprule
Parameter & {Device A (D120)} & {Device B (D122)} \\
\midrule
$a$ & \SI{448}{\nano\meter} & \SI{448}{\nano\meter} \\
$w$ & \SI{92}{\nano\meter}  & \SI{93}{\nano\meter} \\
$r$ & \SI{167}{\nano\meter} & \SI{172}{\nano\meter} \\
$d_0$ & \SI{70}{\nano\meter}     &   \SI{76}{\nano\meter} \\
$h_0$ & \SI{194.5}{\nano\meter}     &   \SI{196.9}{\nano\meter} \\
$d_{17}$ & \SI{122}{\nano\meter}     &   \SI{123}{\nano\meter} \\
$h_{17}$ & \SI{217.6}{\nano\meter}     &   \SI{231}{\nano\meter} \\
$u_y$ & \SI{356}{\nano\meter} & \SI{359}{\nano\meter} \\
fillet radius & \SI{25}{\nano\meter} &   \SI{25}{\nano\meter} \\
$\delta_{x}$ & \SI{4.2}{} &   \SI{3.68}{} \\
$M$ & \SI{2.55}{} &   \SI{2.55}{} \\
\bottomrule
\end{tabular}
\end{table*}

\section{Onset of the strong coupling regime}

\begin{figure*}[htbp]
\centering
\includegraphics{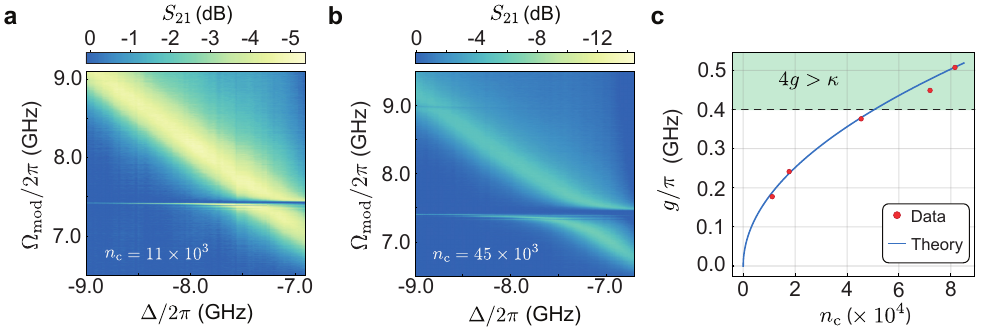}
\caption{\textbf{Onset of the strong coupling regime.} Optomechanically-induced transparency measurement as a function of laser-cavity detuning $\Delta$ for \textbf{a}, $n_\text{c} = 11\times 10^3$ and \textbf{b}, $n_\text{c} = 45 \times 10^3$. \textbf{c}, Mode splitting due to OMIT as a function of $n_\text{c}$. The green-shaded area denotes the optomechanical strong coupling regime. The theory trace was obtained using the measured $g_0$ from dynamical backaction experiments shown in the main text.}
\label{Suppl:OMIT_sc}
\end{figure*}

Additional optomechanically-induced transparency (OMIT) experiments were performed below and above the threshold for the strong coupling regime, as exemplified in~\ref{Suppl:OMIT_sc}a and b. The splitting resulting from OMIT -- a direct measurement of $g/\pi$ -- agrees well with predictions assuming the independently measured $g_0$ from cooling and heating experiments described in the main text, as displayed in~\ref{Suppl:OMIT_sc}c.

\FloatBarrier

\section{Optical stability at \SI{3}{\kelvin}}
\label{section:suppl_optstability}
\begin{figure*}[!htbp]
\centering
\includegraphics{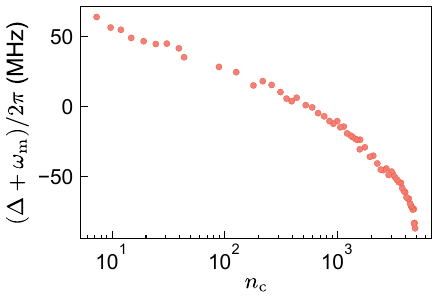}
\caption{\textbf{Optical Power stability.} Frequency shift of the optical mode versus cavity photon occupation $n_\text{c}$ measured using coherent spectroscopy of the optical mode.}
\label{Suppl:optstability}
\end{figure*}

We intensity modulate a pump laser ($\Delta = -\omega_\text{m}$) at a frequency $\Omega_\text{mod}$ using an electro-optical intensity modulator and a vector network analyzer. The reflected signal from the device is measured on a high-speed photodiode. The beating signal between the carrier and sideband contains information about the optical mode of the device.
We track the detuning (\ref{Suppl:optstability}) between the unlocked laser and the location of the optical mode, depending on the incoming laser power.
We observe shifts $<\SI{150}{\mega\hertz}$ even for cavity occupations up to $n_\text{c} \approx 5000$.

\FloatBarrier

\section{Measurement Setup}

\begin{figure*}[htbp]
\centering
\includegraphics[width = 18cm]{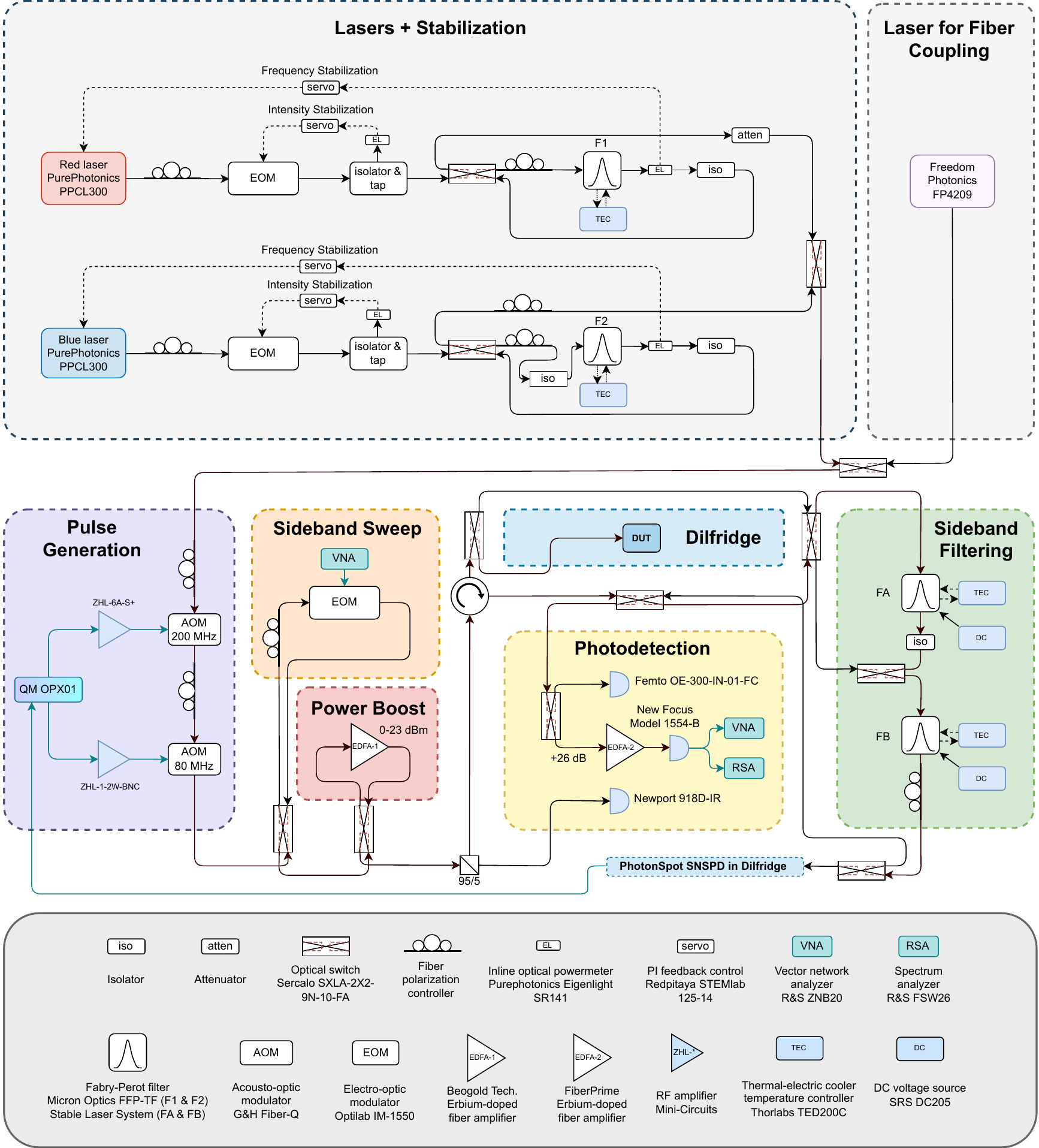}
\caption{\textbf{Measurement Setup.} Diagram representing the full optical setup used in the pulsed and CW measurements.}
\label{Suppl:setup}
\end{figure*}

A schematic of the full experimental setup used in the sideband asymmetry, ground state cooling, and optomechanical strong coupling is provided in~\ref{Suppl:setup}. Two tunable diode lasers (PurePhotonics PPCL300) are first intensity-stabilized with electro-optic modulators (EOMs), and then frequency-stabilized using temperature-stabilized fiber Fabry-P\'{e}rot filters (F1 and F2). A fast wavelength-scanning laser (Freedom Photonics FP4209) assists with fiber-to-chip coupling in the dilution refrigerator. Two acousto-optic modulators (AOMs) in-series are simultaneously pulsed, utilizing a Quantum Machine OPX, to generate the optical pump pulse with a high on-off ratio ($> \SI{90}{\decibel}$) for the pulsed operation. On the other hand, sending a constant signal to the AOMs enables continuous-wave (CW) operation. A pair of narrow-band Fabry-P\'{e}rot cavities (FA and FB), with bandwidth $\approx \SI{15}{\MHz}$, suppress the pump photons with a joint suppression of $>\SI{90}{\dB}$ with respect to the sideband photons. Multiple MEMS optical switches route light to various segments of the setup: an EOM to generate and sweep sidebands (used for OMIT measurements, locking filter cavities FA and FB, calibrating the cooling measurement), an erbium-doped fiber amplifier (EDFA) to boost input power to the device (for the strong-coupling measurement), filter cavities FA and FB to suppress the pump before single photon detection (Photonspot, used for pulsed asymmetry measurements), and a high-speed photodetector (for OMIT, cooling, and strong-coupling measurements).

\FloatBarrier

\end{document}